\documentstyle[twocolumn]{jpsj}

\def\beq{\begin{equation}}
\def\eeq{\end{equation}}
\def\beqa{\begin{eqnarray}}
\def\eeqa{\end{eqnarray}}
\def\beqan{\begin{eqnarray*}}
\def\eeqan{\end{eqnarray*}}
\def\beqp{\begin{equation}\left\{\begin{array}{ll}}
\def\eeqp{\end{array}\right.\end{equation}}
\def\bfg{\begin{figure}}
\def\efg{\end{figure}}
\def\bc{\begin{center}}
\def\ec{\end{center}}

\def\runtitle{Quantum Monte Carlo Study on Magnetization Processes}
\def\runauthor{Hiroaki {\sc Onishi}, Masamichi {\sc Nishino}, Naoki {\sc Kawashima}
and Seiji {\sc Miyashita}.}

\title{\bf Quantum Monte Carlo Study on Magnetization Processes}

\author{Hiroaki {\sc Onishi}, Masamichi {\sc Nishino}$^1$, Naoki {\sc Kawashima}$^2$
and Seiji {\sc Miyashita}}

\inst{Department of Earth and Space Science, Graduate School of Science, Osaka University\\
$^1$Department of Chemistry, Graduate School of Science, Osaka University\\
Machikaneyamacho 1-1, Toyonaka 560-0043\\
$^2$Department of Physics, Tokyo Metropolitan University
    Minami-Ohsawa 1-1, Hachiohji, Tokyo 192-0397}

\recdate{\today}

\abst{
A quantum Monte Carlo method combining update of
the loop algorithm with the global flip of the world line is proposed
as an efficient method for studying the magnetization process in an external field,
which has been difficult because of inefficiency of the update of the total magnetization.
The method is demonstrated in the one dimensional antiferromagnetic Heisenberg model
and the trimer model.
We attempted various other Monte Carlo algorithms to study systems in the external field
and compared their efficiency.
}

\kword{\rm \ Quantum Monte Carlo method,\ Magnetization process,\ Loop algorithm,\ Global flip
}

\begin{document}
\sloppy
\maketitle

\section{Introduction}

The quantum Monte Carlo method has become one of the most
powerful methods to study strongly correlated quantum systems.
Since this method was proposed, various improvements have been introduced.\cite{qmc}
Several problems have been pointed out concerning the efficiency of the method
in the original style which is called  the world line quantum Monte Carlo method (WLQMC),
such as long autocorrelation in Monte Carlo update,
the nuisance of the extrapolation of the Trotter number,
inefficient sampling in the study of the magnetization process in an
external field\cite{roji,magpro}
and the negative sign problem.\cite{negsign}

Recently what is called the loop algorithm has been introduced\cite{loop}
to overcome the problem of long autocorrelation in Monte Carlo update.
Furthermore instead of discrete time with finite Trotter number,
an algorithm using continuous time has been introduced\cite{cont}
and the nuisance of the extrapolation of the Trotter number has been overcome.
These improvements allow us to study systems at very low temperature.

So far some successful approaches to the magnetization process
have been reported.\cite{worm,sandvik}
In this paper we propose a direct application of the loop algorithm
with continuous time method (LCQMC) to the study of the magnetization process.
There are several realizations of LCQMC in the external field.
We attempted various methods and compared their efficiency.
We found that a hybridization of the standard LCQMC and the global flip in WLQMC
is particularly efficient for all values of the field.
We demonstrate this method for the one dimensional antiferromagnetic Heisenberg
model (1DAFH) and also for the trimer model.\cite{trimer1,trimer2}

\section{Model and Method}

In the quantum Monte Carlo method using the Suzuki-Trotter decomposition:
\beq
{\rm e}^{-\beta(A+B)} \cong ({\rm e}^{-{\beta\over N} A}{\rm e}^{-{\beta\over N} B})^N,
\eeq
we express a $d$-dimensional quantum state in a $(d+1)$-dimensional classical
configuration. The new axis is called the Trotter direction which comes from
the decomposition. \cite{qmc}
The thermal average of quantities is obtained by sampling in the classical configuration.

In WLQMC, in order to change the total magnetization,
what is called the global flip is carried out
where a straight world line is flipped simultaneously.
It is, however, very rare that the global flip is accepted
at low temperature in systems with the strong quantum fluctuation,
because the straight world line exists with a very low probability.\cite{magpro}
On the other hand, in LCQMC, we generate various shapes of the loop
which has nonzero magnetization.
The flip of such loops causes change of the total magnetization.
Thus it would be expected that the problem of
inefficient sampling in Monte Carlo study of the magnetization process in an external field
can be solved in LCQMC.

In order to investigate the feasibility of the method,
we study the magnetization process of the one dimensional
antiferromagnetic Heisenberg model:
\beq
{\cal H}=J\sum_{i=1}^L \mbox{\boldmath $S$}_i\cdot\mbox{\boldmath $S$}_{i+1}-H\sum_{i=1}^L S_i^z,
\label{hamafh}
\eeq
where $S_i^{\alpha}={1\over2}\sigma_i^{\alpha}$ ($\sigma^{\alpha}$ is the Pauli
matrix), $\alpha=x,y,$ and $z$.
Here a periodic boundary condition is adopted.
According to the standard loop algorithm,\cite{loop}
a graph is assigned to each plaquette where four spins interact.
The spin configurations of the plaquette and the types of the graph
are depicted in Figs.~1(a) and 1(b), respectively.
In Table I the weights $\{w_{\rm g}({\rm C})\}$ of the graphs \{g=G1,G2,G3,G4\}
in the spin configurations \{C=C1,C2,C3,C4\} for $H=0$ are listed, where
\beq
\begin{array}{ccl}
w_1 &=& {\displaystyle \exp\left(-{\Delta \tau\over 4}J\right) },\\
\rule[0pt]{0pt}{15pt}
w_2 &=& {\displaystyle \exp\left({\Delta \tau\over 4}J\right)\sinh\left({\Delta \tau\over 2}J\right) },
\end{array}
\eeq
there
\beq
\Delta\tau={\beta \over N},
\eeq
with the Trotter number $N$.
$W$(C) is the Boltzmann factor of the configuration C and
\beq
W({\rm C}) = \sum_{\rm g} w_{\rm g}({\rm C}).
\label{weighteq}
\eeq
Graphs are allocated with the weights $\{w_{\rm g}({\rm C})\}$ and they form loops.

In the continuous time algorithm,\cite{cont}
\beq
N\rightarrow \infty, \quad {\rm that \ \ is} \quad {\beta\over N} \rightarrow 0,
\eeq
a horizontal cut which corresponds to G2 in Fig.~1(b) is allocated
with a probability density $J/2$
where the spins in both sides are antiparallel which corresponds to C2 in Fig.~1(a).

The effect of the external field $H$ is taken into account through the global weight.\cite{loop}
That is, we flip a loop with a magnetization $m$ with a probability
\beq
p(m)={\exp(-\beta Hm)\over\exp(\beta Hm)+\exp(-\beta Hm)},
\eeq
where $m$ is defined by
\beq
m = \frac{1}{\beta} \oint_{\rm loop} S^z(\tau){\rm d}\tau.
\eeq
Here $\oint_{\rm loop}$ denotes an integration along a loop
and $m$ takes values $0,\; \pm\frac{1}{2},\; \pm1,\; \cdots$.

The Monte Carlo results for $L=12$ are compared with the exact ones obtained
by the diagonalization method.
It should be noted that the efficiency of Monte Carlo sampling for the magnetization process
becomes better if the length of the chain increases,
because the efficiency depends on how often processes
changing the total magnetization are accepted.
Generally if we can produce a good result for the magnetization process in a short chain,
we expect successful results in longer chains.
Thus in this paper we investigate the efficiency in a short chain $L=12$.
The magnetization processes were obtained
with 1,000,000 Monte Carlo steps (MCS).
The simulation of $10^6$ MCS is divided into 10 bins (each bin has $10^5$ MCS).
The errorbar is estimated from the standard deviation
of the distribution of the data of the bins.
It should be noted that errorbar in this definition becomes very small
if the configuration freezes in some local stable configuration during Monte Carlo simulation,
where all the bins give almost the same value.
The true errorbar should be obtained from the distribution of the data
of several independent simulations.
In this paper, however, the errorbar denotes the one in the former definition
as far as no particular attention is payed.

In Fig.~2(a) we show the magnetization process at a high temperature ($T/J=1.0$),
where the circle denotes the Monte Carlo result and the solid line denotes the exact value.
There we find full agreement.
Furthermore it shows excellent convergence of the data with small error.
Thus the method produces good results as was expected at high temperature.
In Fig.~2(b) we show the magnetization process at a low temperature ($T/J=0.1$).
There the present method reproduces the exact data at low fields up to $H \simeq J$.
At higher fields, however, the method fails to produce the correct magnetization,
where a kind of freezing of the configuration seems to occur.
This observation can be interpreted as follows.
The loops to be flipped are made by assigning graphs with the weight in Table I
which is suitable for $H=0$.
Thus update with the flip of such loops is considered to be an important sampling for $H=0$
and efficient to study low field properties.
On the other hand, this sampling is no more efficient at high field,
although it would provide the correct result with infinite MCS in principle.
The configuration remains in a fixed magnetization
and the method can not produce the correct result.
In order to overcome this difficulty, we have to introduce other recipes.
In the following, we examine two methods to overcome this difficulty.

\subsection{LCQMC using weights for $H\ne0$}

One plausible way to overcome the difficulty is to use graphs for $H \ne 0$.
The weight $w_{\rm g}({\rm C})$ consists of two parts:
\beq
w_{\rm g}({\rm C}) = v_{\rm g}\Delta_{\rm g}({\rm C}),
\label{weightdec}
\eeq
where $\Delta_{\rm g}({\rm C})=1$ or $0$.
The weight for $H=0$ shown in Table I
corresponds to a set of \{$\Delta_{\rm g}$(C)\} shown in Table II and
\beq
\begin{array}{ccl}
v_{{\rm G1}} &=& {\displaystyle 1-{\Delta\tau\over 4}J },\\
\rule[0pt]{0pt}{15pt}
v_{{\rm G2}} &=& {\displaystyle {\Delta\tau\over 2}J },
\end{array}
\eeq
in the first order of $\Delta \tau$.
For $H \ne 0$ the Boltzmann factors of C1 and C1' are different.
In this case there is no solution for the graph weights $\{v_{\rm g}\}$
within $\{\Delta_{\rm g}({\rm C})\}$ given in Table II.
Thus we have to choose another set of $\{\Delta_{\rm g}({\rm C})\}$
and look for a positive solution $\{v_{\rm g}\}$
for the eqs.~(\ref{weightdec}) and (\ref{weighteq}).
For example if we choose a set shown in Table III,
we find a solution
\beq
\begin{array}{ccl}
v_{{\rm G1}} &=& {\displaystyle 1-{\Delta\tau\over4}J-{\Delta\tau\over2}H },\\
\rule[0pt]{0pt}{15pt}
v_{{\rm G2}} &=& {\displaystyle {\Delta\tau\over2}J },\\
\rule[0pt]{0pt}{15pt}
v_{{\rm G3}} &=& {\displaystyle {\Delta\tau\over2}H },\\
\rule[0pt]{0pt}{15pt}
v_{{\rm G4}} &=& {\displaystyle {\Delta\tau\over2}H }.
\end{array}
\eeq
Because the effect of the field is taken into account in the graph weight,
we flip each loop with a probability $1/2$.
However, the new weight violates the detailed balance.
In order to recover the detailed balance, some types of flip must be prohibited.
In the above case we do not flip loops which have either the cross cut G3 or
the freezing cut G4 in the configuration C1.

In Fig.~3(a) we show the magnetization process at a high temperature ($T/J=1.0$),
where the circle denotes the Monte Carlo result
and the solid line denotes the exact value.
There we find full agreement as well as the standard LCQMC.
In Fig.~3(b) we show the magnetization process at a low temperature ($T/J=0.1$).
The data are consistent with the exact value at low fields.
At high fields, however, it seems that a kind of freezing of the configuration occurs
as was observed in the standard LCQMC.
Change of the total magnetization hardly occurs at high field and low temperature,
because the number of loops with prohibited cuts increases.

There are several other choices of $\{\Delta_{\rm g}({\rm C})\}$
where we can find a positive solution $\{v_{\rm g}\}$.
We also examined such choices but we found the similar difficulty in all the cases.

\subsection{Hybridization of LCQMC and the global flip in WLQMC}

In this circumstance we introduce the global flip of the world line,
remembering WLQMC can produce $M(H)$ at high field rather well.
We perform the following procedures.
First we perform one step of the standard LCQMC and then
we look for a straight world line and flip the whole line with a probability 
\beq
p_{\rm gf} = \frac{w(-S^z)}{w(S^z)+w(-S^z)},
\eeq
where $S^z$ is the $z$-component of the spin in the straight line
and $w(S^z)$ is defined as
\beq
\log(w(S^z)) = \sum_j \int_0^{\beta} \log(W_j(\tau)){\rm d}\tau + \beta HS^z,
\eeq
where the summation over $j$ denotes all bonds connected to the concerning site,
i.e., here the left and right nearest neighbors.
When the spin in the neighboring line is parallel or antiparallel,
the weight $W_j(\tau)$ is given by
\beq
\log(W_j(\tau)) = -{J\over 4},
\eeq
or
\beq
\log(W_j(\tau)) =  {J\over 4},
\eeq
respectively.

In Fig.~4(a) we show the magnetization process at a high temperature ($T/J=1.0$),
where the circle denotes the Monte Carlo result and the solid line denotes the exact value.
There we find full agreement as well as the previous methods.
In Fig.~4(b) we show the magnetization process at a low temperature ($T/J=0.1$).
Here we find a good result even at the low temperature.
There the uneven magnetization process due to the finiteness of the chain
is also reproduced correctly with small error.
We find this hybrid method practically useful,
although the method has also the difficulty at much low temperature.

Here it should be noted that if we start from a fully magnetized state,
a straight world line appears as a loop in LCQMC.
Thus one may consider that $M(H)$ can be correctly produced only by LCQMC
if we start from a high field with the fully magnetized state
and then reduce the field gradually.
So far we started simulation for the case of $H=0$ and then continued simulation
increasing the field gradually without reset of the spin configuration.
In Fig. 5 we show results obtained by simulation starting from the high initial field
for both the standard LCQMC and the hybrid method.
The reason of this failure in the standard LCQMC is the following.
In LCQMC, the probability of the flip of a straight world line surrounded by
parallel straight lines is
\beq
p_{\rm LC}= {1\over {\rm e}^{\beta H}+1},
\eeq
while the probability of the flip of the line in the global flip is
\beq
p_{\rm gf}= {1\over {\rm e}^{\beta(H-J)}+1}.
\eeq
Thus we find that
the escape rate from the metastable configuration, i.e., a flip of the straight
world line surrounded by parallel ones, is much larger in the global flip.
Thus the hybrid method gives better result.

\section{Magnetization Process of the Trimer System}

The compound ${\rm 3CuCl}_2\cdot{\rm 2dioxane}$ has been studied as a trimer system.\cite{trimer1}
The magnetization process of the trimer system has attracted interest.\cite{trimer2}
The difficulty of Monte Carlo simulation of WLQMC was pointed out.\cite{roji} That is, 
if we use a large Trotter number, change of the total magnetization
hardly occurs as we mentioned above.
In such situation, if we try to estimate the extrapolated
value from rather small values of the Trotter number, it is quite possible to
conclude an apparent extrapolated value which is very different from the
true one.

Here we apply the hybrid method to this model.
The Hamiltonian of the model is given by
\beqa
\label{hamtri}
{\cal H} &=& \sum_{i=1,{\rm step} 3}^{L-2}
[-J_{\rm F}\mbox{\boldmath $S$}_i\cdot\mbox{\boldmath $S$}_{i+1}
-J_{\rm F}\mbox{\boldmath $S$}_{i+1}\cdot\mbox{\boldmath $S$}_{i+2} \nonumber \\
&&+J_{\rm AF}\mbox{\boldmath $S$}_{i+2}\cdot\mbox{\boldmath $S$}_{i+3}]
-H\sum_{i=1}^L S_i^z.
\eeqa

First we study the magnetization process for
\beq
\gamma = J_{\rm F}/J_{\rm AF}=5,
\eeq
which is close to the ratio for the compound ${\rm 3CuCl}_2\cdot{\rm 2dioxane}$.\cite{trimer1,roji}
In Fig.~6 we show $M(H)$ for $L=12$ at a low temperature ($T/J_{\rm AF}=0.1$),
where the circle denotes the Monte Carlo result and the solid line denotes the exact value.
There we find full agreement.
We also found that we can obtain good results even by the standard LCQMC in the trimer system.
The reason is that the present system contains ferromagnetic bonds
where we allocate the cross cut G3.
This cut causes the loop to prolong along the Trotter axis
and thus the number of loops with nonzero magnetization increases.
The flip of such loops contributes to the equilibration of the magnetization.

%
%

Next we study temperature dependence of $M(H)$ for $\gamma = 4.4$,
which has been estimated for the compound ${\rm 3CuCl}_2\cdot{\rm 2dioxane}$
by Hida ($J_{\rm F}=136$ K, $J_{\rm AF}=30.6$ K).\cite{trimer2}
In Fig.~7 we show the Monte Carlo data of $M(H)$ for $L=60$,
which is considered to be long enough to represent the magnetization
process in the thermodynamic limit.
We find a similar temperature dependence of $M(H)$ to the experimental data.
However, $M(H)$ at low temperatures ($T/J_{\rm AF}=0.049$ and $0.14$)
in the experiment show steeper gradient near the saturated field
than those in Fig.~7.
This disagreement suggests that
the Hamiltonian (\ref{hamtri}) does not represent the experimental situation completely,
although it gives a good approximation of the system.
Attempt at more precise tuning of the parameters to fit the data
would provide further information of the material,
which is an interesting problem in the future.

When we decrease the ratio $\gamma$,
the system shows a $1/3$ plateau in the magnetization process.\cite{trimer2}
In Fig.~8 we show temperature dependence of $M(H)$ for the ratio $\gamma = 1$ for $L=60$.
There we find the plateau at very low temperature ($T/J_{\rm AF}=0.05$).
The step like structure is smoothed out at rather low temperature ($T/J_{\rm AF}=0.1$)
and a smooth curve is obtained at a high temperature ($T/J_{\rm AF}=1.0$).

At low temperature and high field
the autocorrelation of the simulation is large even in
the present improved method and the errorbar defined previously does not
represent the correct variation of the data.
For such cases we performed five independent runs
and estimated the standard deviation of data over the runs,
which indicates how much data scatter.
In Fig.~8 the errorbars for data shown by painted symbols denote
the standard deviation of the data in five independent runs of $10^6$ MCS,
while the errorbars for data shown by open symbols denote
the errorbars in a run of $10^6$ MCS.
At high temperatures ($T/J_{\rm AF}=0.1$ and $1.0$),
the errorbars estimated from independent runs are small,
where the single run is considered to be enough.

\section{Summary and Discussion}

We attempted several methods to find efficient Monte Carlo algorithm
for the magnetization process and found a hybridization of the standard
LCQMC and the global flip in WLQMC to be efficient and practically useful.
We applied the method to
the one dimensional antiferromagnetic Heisenberg model
which is a lattice consisting of only antiferromagnetic bonds,
which is regarded as the hardest model to study the magnetization process
by quantum Monte Carlo methods.
Using the method we also obtained the temperature dependence of the
magnetization process of the trimer system successfully,
which has been difficult in WLQMC.

So far various interesting quantum phases have been proposed in the
ground state. However, usually in experiments only data at finite
temperatures are available. If we see the properties at finite temperature
at zero field, it is rather difficult to find characteristic properties of
the quantum phases. Thus the data at finite field are useful
to grasp the characteristics.
The present method is powerful for such purposes.
In particular the position of the 
cross of the magnetization processes is a point below which the
magnetization decreases as the temperature decreases while 
above which it increases as the temperature decreases.
Thus how the magnetization processes cross each other
may represent a characteristic of the system.
It would be an interesting problem to characterize properties
at finite temperature for various types of quantum phases
even if the phase transition itself is defined only in the ground state.

Among the methods\cite{worm,sandvik} including the present method,
a more detailed study will be required to clarify in what situation
a particular approach is the most efficient.

\section*{Acknowledgements}
The present work is partially supported by Grant-in-Aid from the Ministry of
Education, Science and Culture. They also appreciate for the facility of
Supercomputer Center, Institute for Solid State Physics, University of Tokyo.

\newpage

%
\begin{figure}
\begin{center}
\end{center}
\caption{(a) Spin configurations of the plaquette and
(b) types of the graph}
\label{fig1}
\end{figure}

%
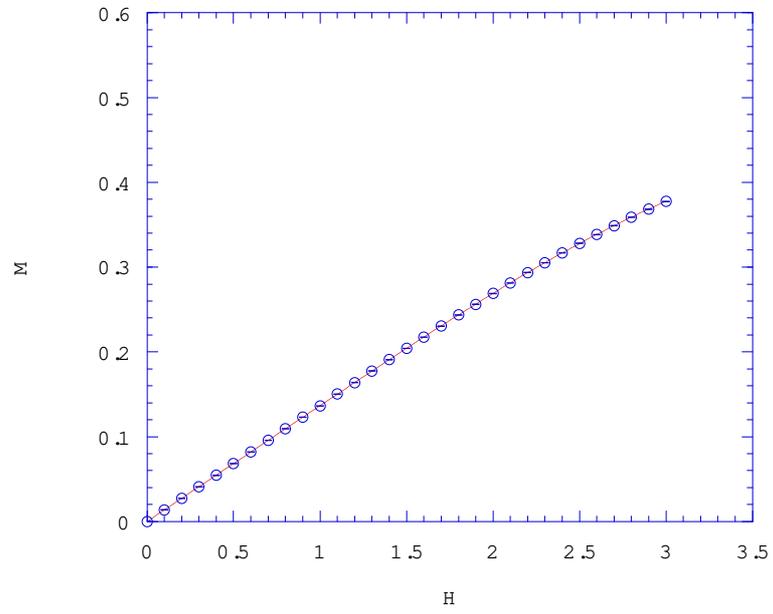
\begin{figure}
\begin{center}
\end{center}
\caption{Magnetization process of the 1DAFH model
obtained by the standard LCQMC with $10^6$ MCS.
(a) $T/J=1.0$ and (b) $T/J=0.1$.
The solid line denotes the exact values obtained by the diagonalization method.
}
\label{fig2}
\end{figure}

%
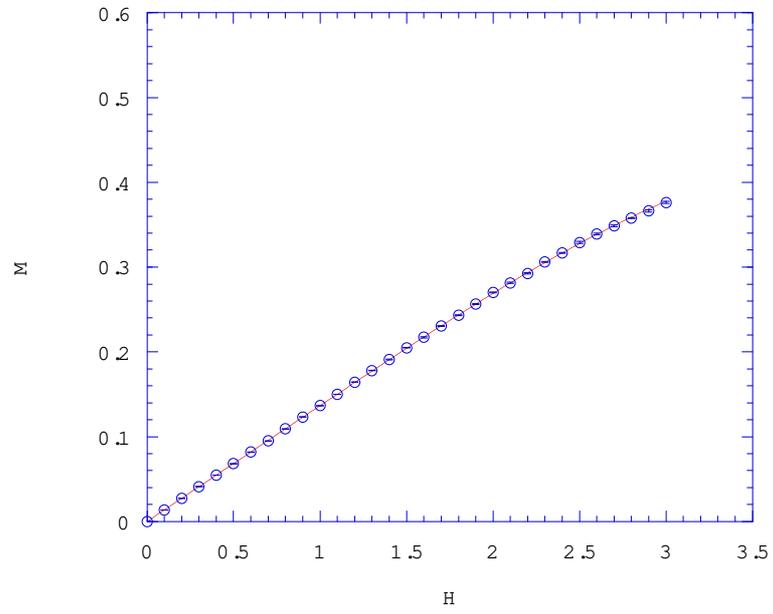
\begin{figure}
\begin{center}
\end{center}
\caption{Magnetization process of the 1DAFH model
obtained by LCQMC with the Table III with $10^6$ MCS.
(a) $T/J=1.0$ and (b) $T/J=0.1$.
The solid line denotes the exact values obtained by the diagonalization method.
}
\label{fig3}
\end{figure}

%
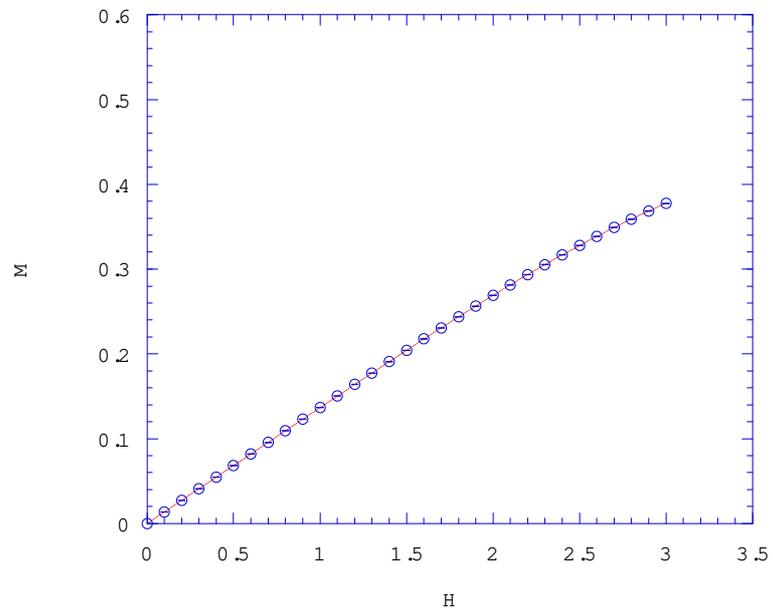
\begin{figure}
\begin{center}
\end{center}
\caption{Magnetization process of the 1DAFH model
obtained by the hybrid method with $10^6$ MCS.
(a) $T/J=1.0$ and (b) $T/J=0.1$.
The solid line denotes the exact values obtained by the diagonalization method.
}
\label{fig4}
\end{figure}

%
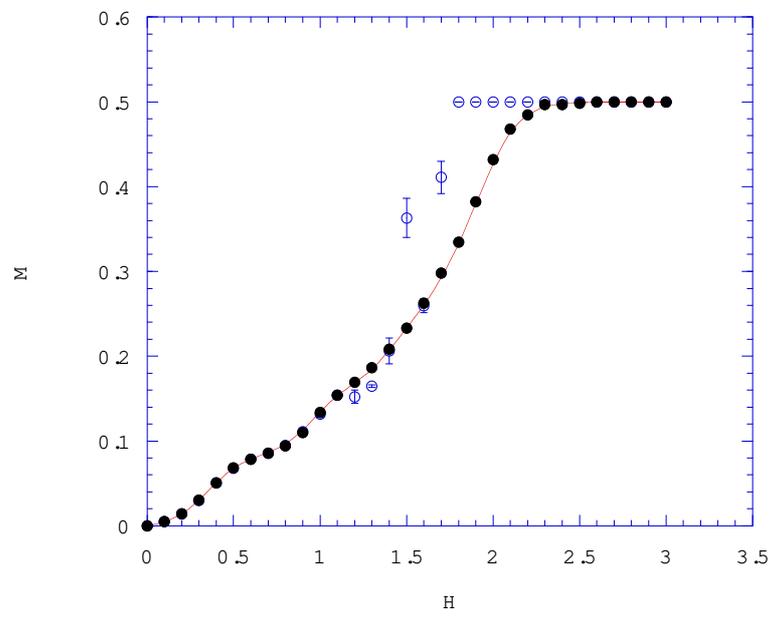
\begin{figure}
\begin{center}
\end{center}
\caption{Magnetization process of the 1DAFH  model
obtained by the standard LCQMC ($\circ$) and the hybrid method ($\bullet$)
starting from a full magnetized state with $10^6$ MCS.
$T/J=0.1$.
The solid line denotes the exact values obtained by the diagonalization method.
}
\label{fig5}
\end{figure}

%
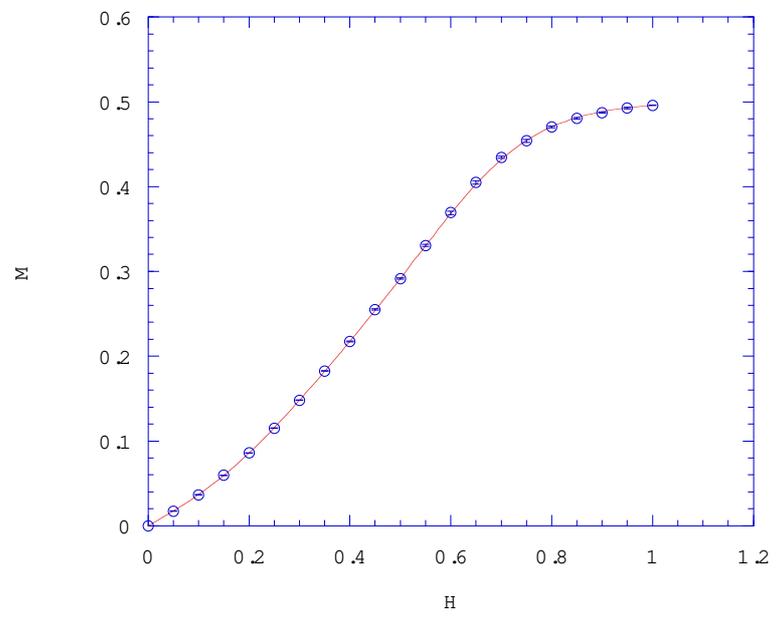
\begin{figure}
\begin{center}
\end{center}
\caption{Magnetization process of the trimer model
obtained by the hybrid method with $10^6$ MCS.
$\gamma=5$ and $T/J_{\rm AF}=0.1$.
The solid line denotes the exact values obtained by the diagonalization method.
}
\label{fig6}
\end{figure}

%
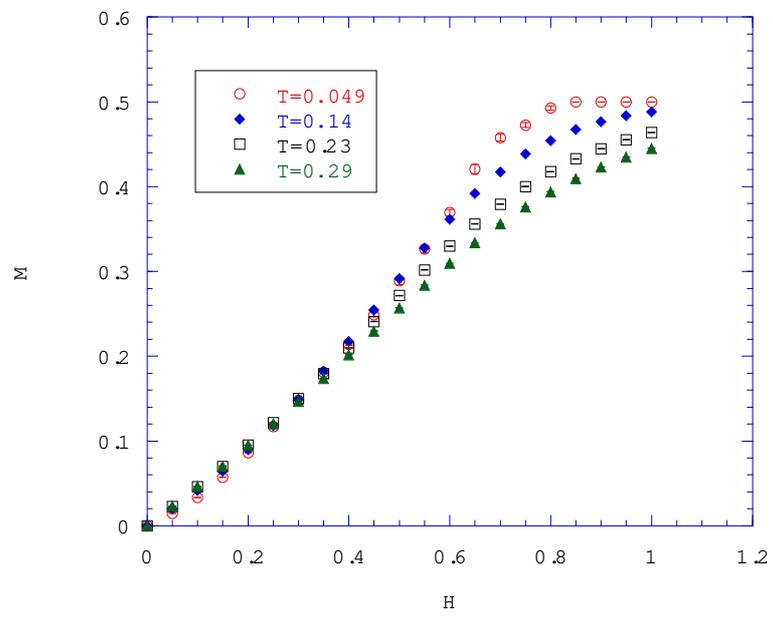
\begin{figure}
\begin{center}
\end{center}
\caption{Magnetization process of the trimer model
obtained by the hybrid method with $10^6$ MCS.
$\gamma=4.4$ and $L=60$.
The symbols circle, diamond, square and triangle denote
$T/J_{\rm AF}=0.049$, $0.14$, $0.23$ and $0.29$, respectively.
These temperatures correspond to $T=1.5$ K, $4.2$ K, $7.0$ K and $9.0$ K, respectively.
The field $H=1.0$ corresponds to $21$ Tesla.
}
\label{fig7}
\end{figure}

%
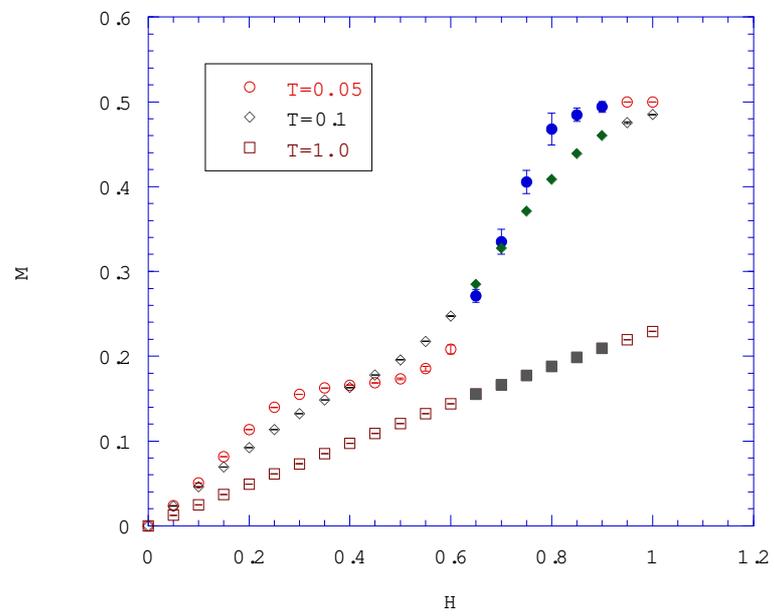
\begin{figure}
\begin{center}
\end{center}
\caption{Magnetization process of the trimer model
obtained by the hybrid method with $10^6$ MCS.
$\gamma=1$ and $L=60$.
The open circle, diamond and square denote
$T/J_{\rm AF}=0.05$, $0.1$ and $1.0$, respectively.
The solid symbols denote the data obtained by five independent runs of $10^6$ MCS.
}
\label{fig8}
\end{figure}

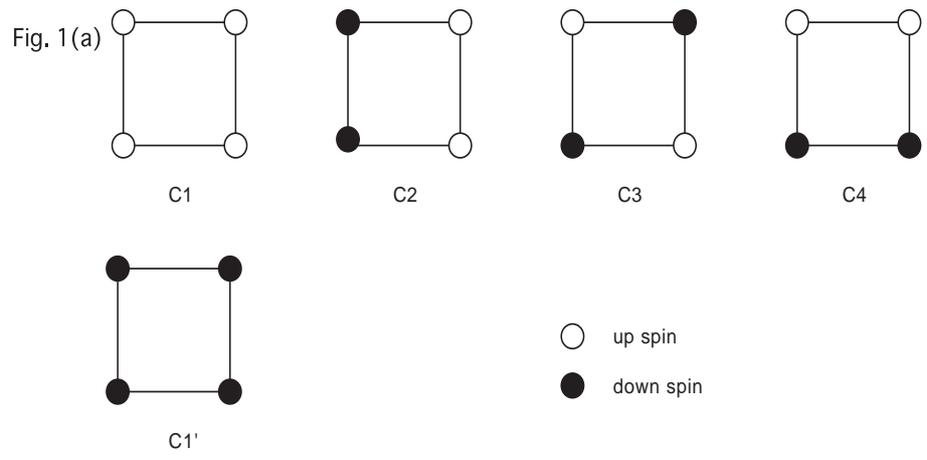
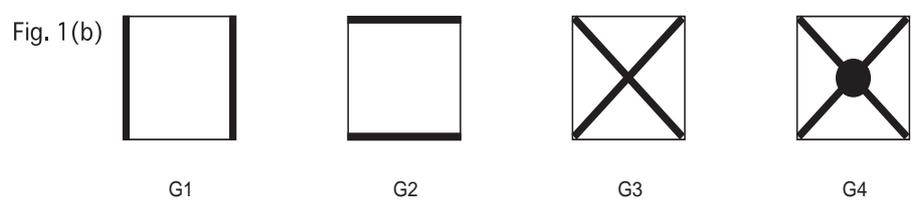
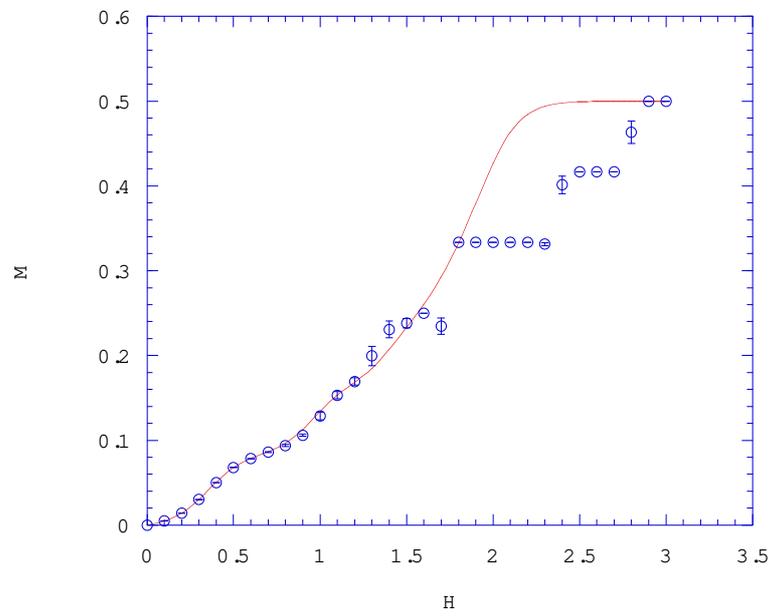
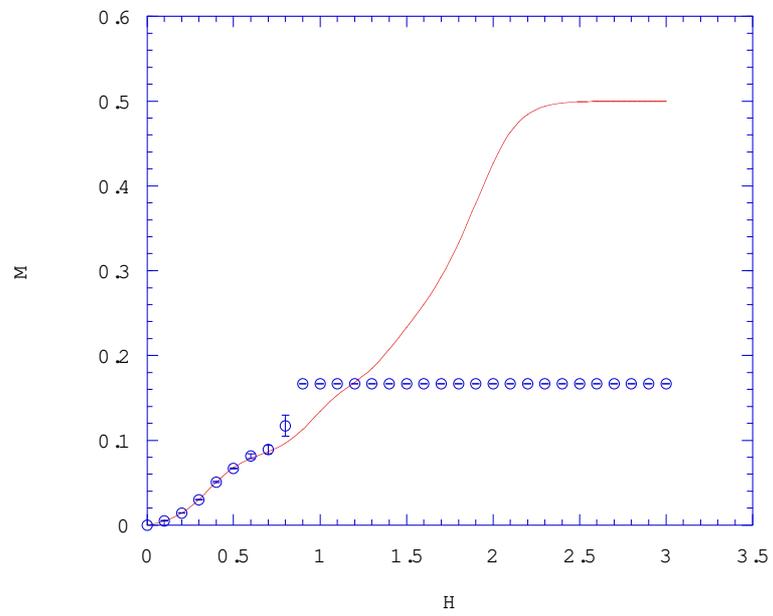
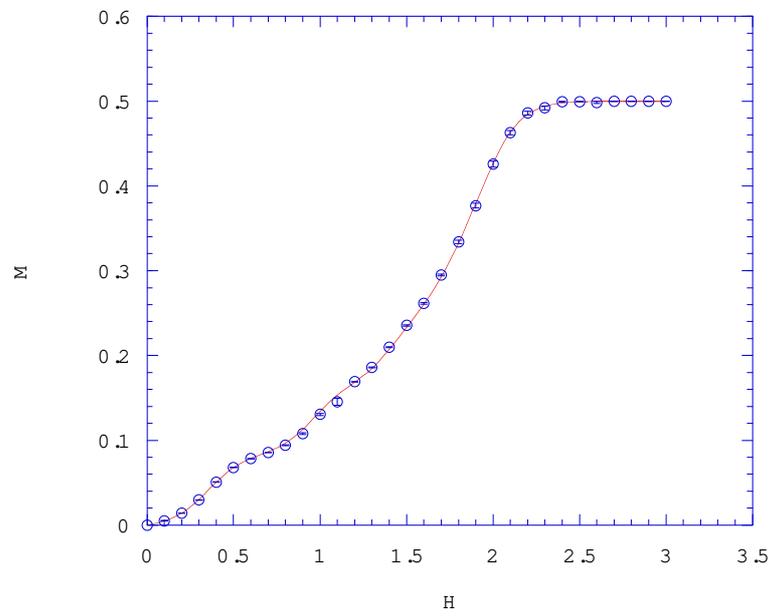
\end{document}